\documentclass[a4paper]{article}

\usepackage{INTERSPEECH2022}
\usepackage{graphicx,xcolor}
\usepackage{multirow}
\usepackage{arydshln}
\usepackage[noend]{algorithmic}
\usepackage{algorithm} 
\newcommand{\ra}[1]{\renewcommand{\arraystretch}{#1}}

\title{Streaming Speaker-Attributed ASR with Token-Level Speaker Embeddings}
 
 \name{Naoyuki Kanda$^1$, Jian Wu$^1$, Yu Wu$^2$, Xiong Xiao$^1$, Zhong Meng$^1$, Xiaofei Wang$^1$, Yashesh Gaur$^1$,\\Zhuo Chen$^1$, Jinyu Li$^1$, Takuya Yoshioka$^1$}
 \address{
   $^1$Microsoft Cloud+AI, USA \hspace{3mm}$^2$Microsoft Research Asia, China} %\\
\email{\{nakanda,wujian,yuwu1,xioxiao,zhme,xiaofewa,yagaur,zhuc,jinyli,tayoshio\}@microsoft.com}
\begin{document}

\maketitle
\begin{abstract}
This paper presents a streaming speaker-attributed automatic speech recognition (SA-ASR) model that can recognize ``who spoke what'' with low latency even when multiple people are speaking simultaneously. Our model is based on token-level serialized output training (t-SOT) which was recently proposed to transcribe multi-talker speech in a streaming fashion. To further recognize speaker identities, we propose an encoder-decoder based speaker embedding extractor that can estimate a speaker representation for each recognized token not only from non-overlapping speech but also from overlapping speech. The proposed speaker embedding, named t-vector, is extracted synchronously with the t-SOT ASR model, enabling joint execution of speaker identification (SID) or speaker diarization (SD) with the multi-talker transcription with low latency. We evaluate the proposed model for a joint task of ASR and SID/SD by using LibriSpeechMix and LibriCSS corpora. The proposed model achieves substantially better accuracy than a prior streaming model and shows comparable or sometimes even superior results to the state-of-the-art offline SA-ASR model.
\end{abstract}
\noindent\textbf{Index Terms}: multi-talker speech recognition,
serialized output training, speaker identification, speaker diarization

\section{Introduction}

Speaker-attributed automatic speech recognition (SA-ASR), a task to 
recognize ``who spoke what'' from audio input, has long been studied
for conversation analysis \cite{janin2003icsi,carletta2005ami,fiscus2007rich,barker2018fifth,watanabe2020chime}.
An SA-ASR system typically consists of multiple
modules such as 
speech separation for handling overlapping speech\cite{chen2021continuous}, 
speaker identification (SID) \cite{bai2021speaker} or speaker diarization (SD) \cite{park2021review}
for estimating the speaker identity, and ASR for transcribing each utterance~\cite{yoshioka2019advances}.
While substantial advancement has been made for SA-ASR systems,
such a modular system tends to be complicated, and difficult 
to optimize for the best accuracy.

To further improve the SA-ASR accuracy, 
various attempts have been made to jointly optimize multiple modules %for SA-ASR
either with partially joint approaches (e.g. speech separation and ASR \cite{yu2017recognizing,seki2018purely,chang2019end,chang2019mimo,kanda2019acoustic,kanda2019auxiliary})
or fully joint approaches \cite{el2019joint,kanda2019simultaneous,mao2020speech}.
In particular, the end-to-end (E2E) SA-ASR model of \cite{kanda2020joint} which
jointly performs multi-talker ASR and SID was shown to considerably outperform 
the modular-approach-based systems \cite{kanda2021comparative}.
However, the E2E SA-ASR model is based on the attention encoder-decoder architecture \cite{bahdanau2014neural,chorowski2014end,chorowski2015attention} 
with serialized output training (SOT)-based ASR \cite{kanda2020sot},
which renders the model usable only for offline (i.e. non-streaming) inference.

Only a limited number of studies addressed
the streaming SA-ASR problem through the joint optimization approach.
% the SA-ASR problem with streaming inference.
Shafey et al. \cite{el2019joint} proposed 
to insert speaker role tags (e.g., ``doctor'' and 
``patient'') in output transcriptions for the 
offline recurrent neural network transducer (RNNT)-based ASR.
Saltau et al. \cite{soltau2021understanding} later showed 
a good performance with this idea 
for medical conversation transcription with streaming RNNT models.
However, 
their model is difficult to extend to general scenarios with arbitrary number of speakers
since it requires each speaker to have a unique role.
Their model also cannot deal with overlapping speech
which is
 ubiquitous in human conversations \cite{ccetin2006analysis}.
Lu et al. \cite{lu2021surit} proposed
streaming unmixing, recognition and identification transducer (SURIT)
that jointly performs multi-talker ASR and SID in a streaming fashion.
Their model has two output branches for ASR and two additional 
output branches for SID to generate ASR and SID results for
up to two simultaneous speakers with limited latency.
However, their word error rate (WER) and speaker error rate (SER) were 
significantly worse than the state-of-the-art (SOTA) results 
of the offline SA-ASR models \cite{kanda2021end}.

To address these limitations,
we propose % speaker-aware transformer transducer (SA-TT), 
a novel streaming SA-ASR model
that works with low latency even for overlapping speech.
Our model is based on token-level serialized output training (t-SOT) \cite{kanda22arxiv}, which
 was recently proposed for low-latency multi-talker speech transcription.
To further estimate the speaker identities,
we propose an encoder-decoder based speaker embedding extractor that
can estimate a speaker representation for each recognized token not only from non-overlapping speech but also from overlapping speech.
%even from the overlapping speech.
The proposed speaker embedding, named t-vector, is extracted synchronously with the t-SOT ASR to jointly perform SID/SD and the multi-talker transcription
for arbitrary number of speakers with low latency.
In our evaluation using LibriSpeechMix \cite{kanda2020joint} and LibriCSS \cite{chen2020continuous} corpora,
we show that the proposed model achieves substantially better
accuracy than the prior streaming SA-ASR model. 
The model also yields on par or sometimes even better accuracy than the SOTA offline SA-ASR model.

\begin{figure}[t]
  \centering
  \includegraphics[width=1.0\linewidth]{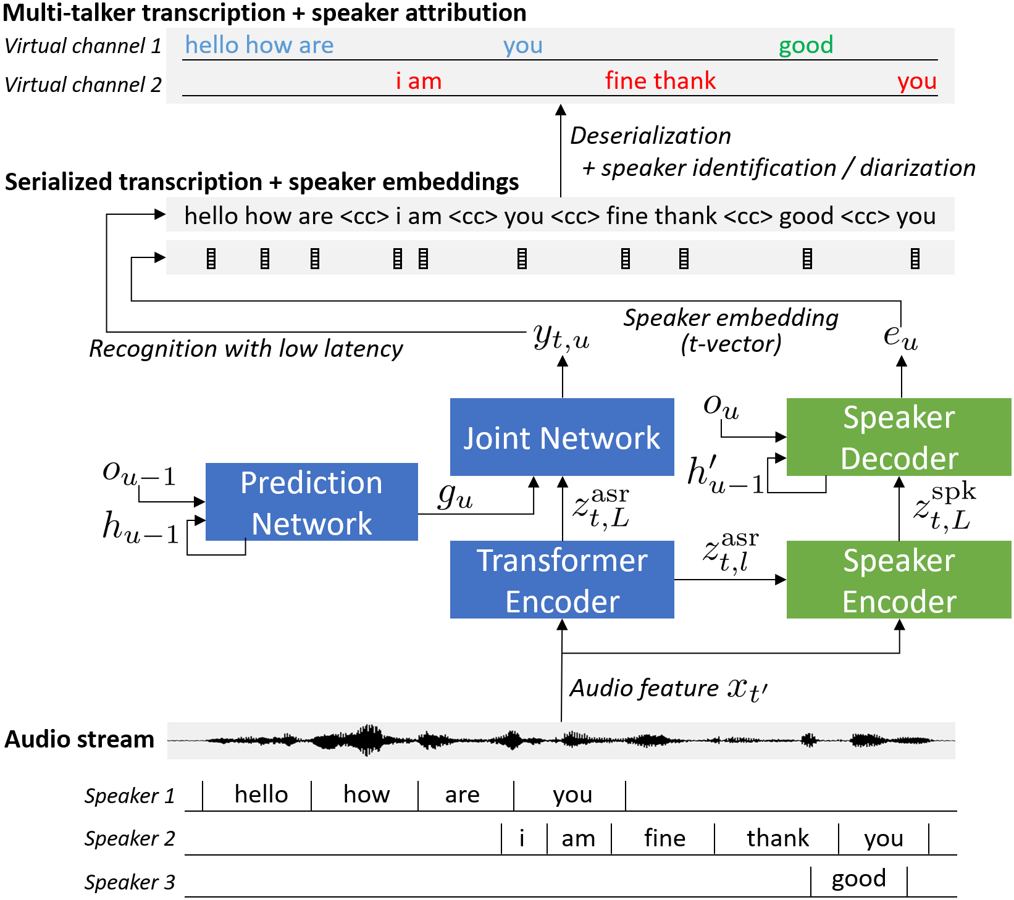}
  \vspace{-7mm}
      \caption{The overview of t-SOT-based TT with t-vector extraction. ASR-related blocks are colored by blue while speaker-related blocks are colored by green.}
  \label{fig:overview}
   \vspace{-5mm}
\end{figure}

\section{Streaming Multi-Talker ASR with t-SOT}
\subsection{t-SOT}

The t-SOT framework was recently proposed to recognize multi-talker conversations with low latency \cite{kanda22arxiv}.
In t-SOT,
we assume up to $M$ utterances are overlapping at the same time in the input audio.
Here, we explain the t-SOT for $M=2$. Refer \cite{kanda22arxiv} for a case with $M>2$.
% An input audio may include any number of utterances as long as the number of concurrent utterances is less than $M$.
In t-SOT,
the transcriptions for multiple speakers are serialized into a single sequence of recognition tokens (e.g., words, subwords) by sorting the tokens in a chronological order. 
A special token $\langle \mathrm{cc}\rangle$, which indicates a change of ``virtual'' output
channels, 
is inserted between two adjacent words if they are spoken by two different speakers.
An example of such a serialized transcription is shown at the middle of Fig. \ref{fig:overview}.
A streaming end-to-end ASR model is trained to generate such serialized transcriptions from the corresponding audio samples.
During inference, a serialized transcription including 
$\langle \mathrm{cc}\rangle$ is produced by the ASR model in a streaming fashion, which is then deserialized to separate transcriptions 
by switching the virtual output channel when $\langle \mathrm{cc}\rangle$ is encountered.
The deseriaization process is illustrated at the top of Fig. \ref{fig:overview}. 
% The t-SOT model was found to outperform prior multi-talker ASR models by a significant margin 
% Refer \cite{kanda22arxiv} for a case with $M>2$.
The t-SOT model was found to significantly outperform prior multi-talker ASR models in both the recognition accuracy and latency 
while keeping
the model architecture as simple as conventional single-talker ASR models \cite{kanda22arxiv}.
% In this paper, we assume $M=2$ for all experiments.

\subsection{Transformer Transducer}

We use transformer transducer (TT) \cite{zhang2020transformer} as a backbone ASR model in this work.
The components of TT are shown by blue blocks in Fig. \ref{fig:overview}.
% Here, we explain an overview of TT 
% to introduce necessary notions for explaining the proposed speaker embedding extractor in the next section.
TT is an RNNT model using Transformer \cite{vaswani2017attention} as the encoder.
% TT consists of transformer-based encoder and recurrent neural network transducer (RNNT)-based decoder.
The TT encoder is represented as follows.
\begin{align}
\hspace{-3mm}z_{[1:T],0}^{asr}&=\mathrm{CNN}(x_{[1:T']}) \\
\bar{z}^{\rm asr}_{t,l}&=z^{\rm asr}_{t,l-1} + \mathrm{MHA}_{\rm l}^{\rm asr}(z^{\rm asr}_{t,l-1},z_{m(t),l-1}^{\rm asr},z_{m(t),l-1}^{\rm asr}), \label{eq:asr-self}\\
z^{\rm asr}_{t,l}&=
\bar{z}^{\rm asr}_{t,l}+\mathrm{FF}_l^{\rm asr}({\bar{z}}^{\rm asr}_{t,l})
 \label{eq:asr-ff}
\end{align}
Here, 
$x_{[1:T']}$ is a sequence of input audio features with length $T'$.
$\mathrm{CNN}()$ represents a stack of convolution layers that subsamples
the sequence length to $T$.
$\mathrm{MHA}^*_l(Q,K,V)$ represents a multi-head attention (MHA) of the $l$-th layer
\cite{vaswani2017attention} with query $Q$, key $K$, and value $V$ matrices.
The streaming mask function $m(t)$ generates limited time frames $[t_s:t_e] \in [1:T]$ to compute the attention at time frame $t$ by using an attention mask.
With an appropriate attention mask,
TT can be executed with limited algorithmic latency as proposed in \cite{chen2021developing}.
Finally, $\mathrm{FF}^{\rm *}_l()$ represents a position-wise feed forward network at the $l$-th layer.
% Here, $\mathrm{Embed}()$ and $\mathrm{PosEnc}()$ are the embedding
% function and absolute positional encoding function \cite{vaswani2017attention}, respectively.
% $\mathrm{MHA}^*_l(Q,K,V)$ represents the multi-head attention of the $l$-th layer
% \cite{vaswani2017attention} with query $Q$, key $K$, and value $V$ matrices.
% $\rm FF_l^{\rm asr}()$ is a position-wise feed forward network in the $l$-th layer.
The decoder of TT consists of the prediction network and joint network as follows.
\begin{align}
g_u, h_u&=\mathrm{Prediction}(o_{u-1},h_{u-1}) \\
% g_u&=\mathrm{Prediction}(o_{[1:u-1]}) \\
y_{t,u}&=\mathrm{Joint}(z^{\rm asr}_{t,L}, g_u) 
\end{align}
Here, $o_{u}$ is the $u$-th estimated token,
and $h_{u}$ is a hidden representation of the prediction network at step $u$.
$L$ is the number of layers in the TT encoder.
The resultant $y_{t,u}$ represents a probability distribution of the tokens at time $t$ given $o_{[1:u-1]}$.
In inference, the token $o_{u}$ is estimated by performing time-synchronous beam search on $y_{t,u}$.

\begin{figure}[t]
  \centering
  \includegraphics[width=1.0\linewidth]{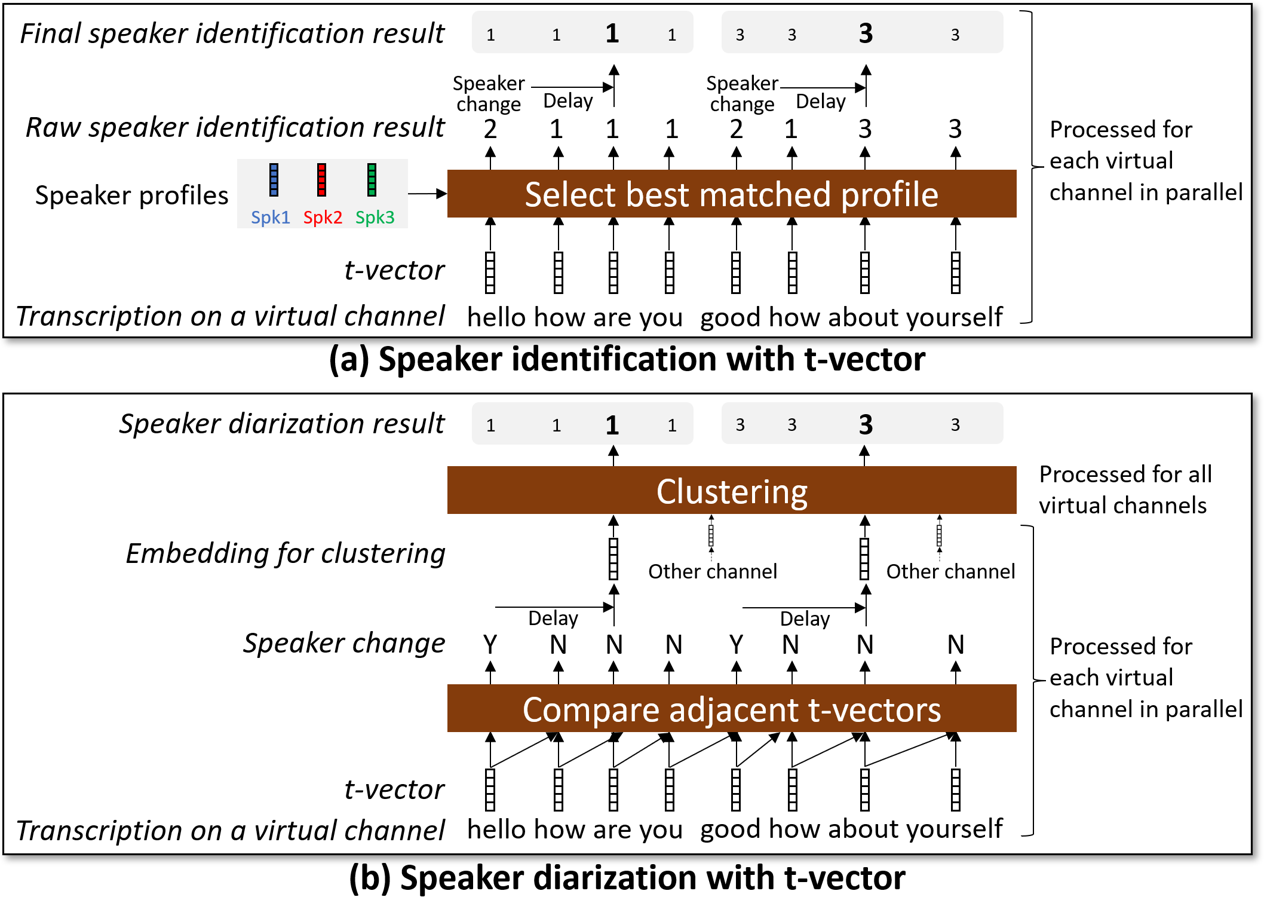}
  \vspace{-7mm}
      \caption{Streaming SID/SD with the proposed model.}
  \label{fig:inference}
   \vspace{-5mm}
\end{figure}

\begin{table*}[t]
\ra{1.0}
  \caption{%SER (\%), WER (\%), and SAWER (\%) 
  Evaluation result for LibriSpeechMix with ASR+SID task
  given 8 speaker profiles,
%    The number of speaker profiles per test audio was 8.
each of which was extracted from 2 utterances. % (15 sec on average). 
% Note that the latency on the ASR is algorithmic latency while the latency on the SID is the
% Average duration on the SID latency is calculated based on the reference word alignment.
The average duration for SID latency is calculated based on the reference alignment.
  }
  \label{tab:librispeechmix_summary}
  \vspace{-3mm}
  \centering
 { \scriptsize
% {  \footnotesize
  % \tabcolsep = 2.0mm
%\begin{tabular}{lcccc}
\begin{tabular}{@{}llrrrrrrrrrr@{}}
    \toprule
Model name & Model type & \multirow{2}{*}{\shortstack[l]{\# of\\param.}} & \multirow{2}{*}{\shortstack[c]{Latency\\on ASR}} & \multirow{2}{*}{\shortstack[c]{ Latency on SID\\(Avg. duration)}}& \multicolumn{3}{c}{1-speaker} &&\multicolumn{3}{c}{2-speaker-mixed}  \\ \cmidrule{6-8} \cmidrule{10-12}
   &  & & &   & SER & WER & {\bf SAWER} && SER & WER & {\bf SAWER}  \\ \midrule % \hdashline[0pt/2pt]\hdashline[0pt/2pt] \midrule
% {\it (Non-streaming Models)}&  \\      
SOT LSTM SA-ASR \cite{kanda2020joint}& Non-streaming  & 146M & $\infty$ & $\infty$ &  0.2 & 4.2 & {\bf 4.5} && 2.5 & 8.7 & {\bf 9.9}\\ 
%%% MEMO: /blob/sdrg/user_std_scus/nakanda/tmp/count_params.py /blob/sdrg/user_std_scus/nakanda/models/spkmix.sot.master.v1.1/sa-asr.conformer-librispeech.3spk.tied.seed320kSA.lr1e-4.warm10k.pred2.no-detach.sw1.0.v5.fix.sc4.G8.1/10000_model.pth.tar
SOT Confomer SA-ASR \cite{kanda2021end}& Non-streaming  & 142M & $\infty$ &  $\infty$ & 0.6 & 3.3 & {\bf 3.9} && 2.0 & 4.3 & {\bf 6.4}  \\ \midrule
% {\it (Streaming Models)}&  \\      
SURIT \cite{lu2021surit} & Streaming & 90M & 0.15 sec & $0.85\mathcal{T}^\dagger$(8.25 sec)  & - & - & - && 6.7$^{\ddagger}$\hspace{-1.5mm} & 10.1 & -  \\ \hdashline[1pt/2pt]\hdashline[0pt/1pt]
t-SOT TT-18 w/ t-vector & Streaming& 100M & 0.16 sec  & + 2 words (0.68 sec) & 2.2  & 5.5  & {\bf 5.9} && 3.8 & 7.9 & {\bf 8.5}   \\ 
t-SOT TT-18 w/ t-vector & Streaming& 100M & 0.16 sec  & + 4 words (1.36 sec) & 1.0 & 5.2  & {\bf 5.5} && 2.0 & 7.3 & {\bf 8.0}   \\ 
t-SOT TT-18 w/ t-vector & Streaming& 100M & 0.16 sec  & + 8 words (2.72 sec) & 0.6 & 5.0  & {\bf 5.4} && 1.2 & 7.0 & {\bf 7.8}   \\ % \hdashline[1pt/2pt]\hdashline[0pt/1pt] 
t-SOT TT-36 w/ t-vector & Streaming& 160M & 0.16 sec  & + 8 words (2.72 sec) & 0.7 & 4.4 & {\bf 4.9} && 1.5 & 6.5 & {\bf 7.4}   \\ 
t-SOT TT-36 w/ t-vector & Streaming& 160M & 2.56 sec  & + 8 words (2.72 sec) & 0.7 & 3.3 & {\bf 4.0} && 1.5 & 4.7 & {\bf 5.8}   \\ \bottomrule 
  \end{tabular}
  }
 \\{\scriptsize %\footnotesize % $^\dagger$ Average audio length was 9.7 sec. 
 % 0.3 sec, 310 msec
 $^\dagger$ $\mathcal{T}$ is the length of the audio segment whose average duration was 9.70 sec for the test set. \;\;$^\ddagger$  Each speaker profile was extracted by using 10 utterances. % (75 sec on average).
 }
\vspace{-6mm}
\end{table*}

% \section{SA-TT: SID / SD with t-SOT TT}
\section{Streaming SA-ASR with t-vector}

\subsection{t-vector: Token-level speaker embedding}
\label{sec:speaker_embedding}

The key challenge of SID/SD with streaming multi-talker ASR is
estimating clean speaker representations {\it even from overlapping speech}
in a way synchronous with 
the ASR output tokens {\it whose emission times can have arbitrary delays} 
from their actual spoken times in the input audio.
To overcome this challenge,
we propose an encoder-decoder based speaker embedding extraction method 
that works synchronously with the emission of token $o_u$ from TT.

The proposed speaker embedding extraction model consists of a speaker encoder
and a speaker decoder as shown by the green blocks
in Fig. \ref{fig:overview}. 
The speaker encoder 
has the same number of layers (i.e. $L$ layers) with the transformer encoder of TT.
It
cooperatively works with the TT encoder and generates
raw speaker representation 
$z^{\rm spk}_{t,L}$ from the final (i.e. $L$-th) layer for each time frame $t$. %a token $o_u$ emitted at time frame $t_u$.
Specifically, the proposed speaker encoder is formulated as follows.
\begin{align}
\hspace{-3mm}z_{[1:T],0}^{\rm spk}&=\mathrm{DvecExtractorNet}(x_{[1:T']}) \\
z^{\rm spk}_{t,l}&=z^{\rm spk}_{t,l-1} + \mathrm{MHA}_{\rm l}^{\rm spk}(z^{\rm asr}_{t,l-1},z_{m(t),l-1}^{\rm asr},z_{m(t),l-1}^{\rm spk}). \label{eq:asr-self}
% z^{\rm spk}_{t,l}&=z^{\rm spk}_{t,l-1}  \nonumber \\ 
%                  & \hspace{-3mm}+ \mathrm{MHA}_{\rm l}^{\rm spk}(z^{\rm asr}_{t,l-1},z_{m(t),l-1}^{\rm asr},z_{m(t),l-1}^{\rm spk}), \label{eq:asr-self}
\end{align}
Here, $\mathrm{DvecExtractorNet}()$ is a neural network with almost the same architecture as a d-vector extractor \cite{variani2014deep} (Res2Net \cite{gao2019res2net,xiao2021microsoft}
in our implementation)
% used for speaker profile extraction in SID task
except that
all look-ahead weights in the convolution layers are zeroed out to prevent the speaker encoder from adding further latency. 
It generates the speaker representation $z_{[1:T],0}^{\rm spk}$ at the bottom (i.e. 0-th) layer. %, which has the same length with $z_{[1:T],0}^{asr}$.
For each layer $l$, MHA is applied by using the query $z^{\rm asr}_{t,l-1}$ and key $z_{m(t),l-1}^{\rm asr}$
from TT while using the value $z_{m(t),l-1}^{\rm spk}$ inside the speaker encoder.
 The masking function $m()$ is shared with TT such that the speaker encoder works with limited latency. % as same with TT.
% Note that
% the projection matrices inside MHA \cite{} 
% is not shared with TT such that 
% an optimal attention distribution for speaker embedding extraction is learnt during the training.

The speaker decoder generates speaker embedding $e_u$, named t-vector,
for each non-blank token, i.e., $o_u$ as follows.
%,including $\langle \mathrm{cc}\rangle$, as follows.
\begin{align}
e_u,h'_{u} % &= \mathrm{SpeakerDecoder}(z^{\rm spk}_{t_u,L}, o_u, h'_{u-1}), \\
    &= \mathrm{LSTM}(z^{\rm spk}_{t_u,L} + \mathrm{Embed}(o_u), h'_{u-1}), \label{eq:spk_dec}
\end{align}
where
 $t_u$ is the emission time frame of $o_u$ based on TT,
 %(i.e. not the reference emission time),
$\mathrm{Embed}()$ is the embedding function, and $h'_{u}$ is the hidden state of 
the long short-term memory (LSTM) at step $u$.
%In training, the value of $t_u$
%is determined by Viterbi time-alignment 
%of the reference tokens based on TT.
%  $t_u \in\{1,...,T\}$ is the emission time frame of $o_u$.
% The resultant $e_u$ can be seen as the speaker embedding corresponding to the token $o_u$.
% Note that Eq. \eqref{eq:spk_dec} works even for
%  $\langle \mathrm{cc}\rangle$ to update $h'_u$ 
%  because
%  $\langle \mathrm{cc}\rangle$ is a strong clue to estimate different embeddings
%  after it.

For training the speaker encoder and decoder, 
we use a simple softmax loss based on speaker identification using the cosine similarity for $e_u$ as
\begin{align}
 \mathcal{L}^{\rm spk}=\hspace{-2mm}\sum_{u|_{o_u\neq \langle \mathrm{cc}\rangle}}{\hspace{-3mm}-\log\frac{e^{cos(e_u, d_u)}}{e^{cos(e_u, d_u)}+\sum_{d'\in \Phi(\mathcal{D})} e^{cos(e_u, d')}}}, \label{eq:spk_loss}
\end{align}
where $d_u$ is a reference d-vector for the speaker of token $o_u$.
$\Phi(\mathcal{D})$ is a random subset of
the d-vectors $\mathcal{D}$ in the training data, where $\Phi(\mathcal{D})$ does not include
the d-vector of the speaker of $o_u$.
In training, the value of $t_u$, which is necessary to 
extract $e_u$, is determined by Viterbi time-alignment 
of the reference tokens based on TT.
Note that 
% the speaker decoder (Eq. \eqref{eq:spk_dec}) works for all tokens 
% including  $\langle \mathrm{cc}\rangle$ to update $h'_u$ because
% $\langle \mathrm{cc}\rangle$ is a strong clue to estimate different embeddings
% for words before and after it.
% On the other hand,
 $\mathcal{L}^{\rm spk}$
%  \eqref{eq:spk_loss}) 
 is computed only for 
 normal tokens except for $\langle \mathrm{cc}\rangle$
 because $e_u$ for $\langle \mathrm{cc}\rangle$ is not used for inference.

\subsection{Streaming ASR+SID with t-vector}
\label{sec:streaming-sid}

In the context of SA-ASR, 
SID is a task to determine the speaker of each word
given a set of speaker profiles (i.e. d-vectors) of candidate speakers. %, i.e. speaker embeddings for candidate speakers.
The proposed procedure is shown in the upper diagram of Fig. \ref{fig:inference}.
Here,
for each virtual output channel,
a t-vector for each word\footnote{When one word consists of
multiple subwords, the t-vector corresponding to the last subword in the word is used.}
% $e_u$ 
is compared with each speaker profile based on the cosine similarity,
and the index of the speaker profile that has the largest cosine similarity
% to $e_u$ 
is selected
as a raw SID result of the word. 
A speaker change is then detected 
based on the change of the raw SID result.
% based on the raw SID result.
A final SID result is determined after a delay of a fixed number of words 
(e.g., 2 words in Fig. \ref{fig:inference}) from
the speaker change point.
% since it usually takes a few iteration of 
% speaker decoder to reliably estimate the speaker representation. 
Note that, once a speaker change is detected,
the next speaker change is not detected until the final SID result is determined
as exemplified in Fig. \ref{fig:inference}
where speaker change is not detected at ``how'' in the second segment.

  The delay parameter mentioned above is introduced 
because it usually takes a few iterations for 
 the speaker decoder to reliably estimate the speaker representation. 
% Generally, the longer the delay is, the better the estimated embedding becomes.
On the other hand, 
the longer the delay is, the higher the chance of missing speaker change points becomes.
 The impact of the delay  will be discussed in the experiment section.

\subsection{Streaming ASR+SD with t-vector}

In the context of SA-ASR, 
SD is a task to assign a speaker cluster index for each spoken word
without having speaker profiles.
The proposed procedure is shown in the lower diagram of Fig. \ref{fig:inference}.
Here, 
for each virtual output channel,
we first detect the speaker change point if
the cosine similarity of 
 adjacent t-vectors is smaller than a pre-determined threshold. 
The t-vector
after a fixed number of words from the speaker change point is used
as the speaker embedding for the segment until 
the next speaker change
is detected.
Every time  a new speaker embedding is generated,
a clustering procedure
%(in our case, normalized maximum eigengap-based
%spectral clustering (NME-SC) \cite{park2019auto}) 
is applied to the speaker embeddings estimated for all the output channels up to the current time. 
This allows
an intermediate SD result to be presented with low latency, 
% By applying the clustering every time speaker embedding is generated,
%A user will perceive the intermediate
%SD result with limited latency,
where the algorithmic latency is determined by the delay in finalizing the speaker embedding for the segment.\footnote{An idea to repeatedly apply clustering was 
proposed in \cite{dimitriadis2017developing}.}
% Note that, in our experiment in the next section, 
% we show only the result at the end of recordings as is done in
% prior works for streaming SD \cite{zhang2019fully,xia2021turn}.

\begin{table*}[t]
\ra{0.93}
% \ra{1.0}
  \caption{cpWER (\%) for LibriCSS with ASR+SID and ASR+SD tasks. Algorithmic latency is shown in the latency columns. 0L and 0S are 0\% overlap conditions with long and short inter-utterance silences.}
%   8 speaker profiles corresponding to the speakers in the recordings were provided for the ASR+SID task. 
%   The oracle number of speakers was used for the ASR+SD task. }
  %The number of speaker profiles per test audio was 8.
%Each speaker profile was extracted by using 5 utterances (45 sec on average). 
%  }
  \label{tab:libricss_summary}
  \vspace{-3mm}
  \centering
 { \scriptsize
%  {  \footnotesize
  % \tabcolsep = 2.0mm
%\begin{tabular}{lcccc}
\begin{tabular}{@{}lrrrrrrrrr@{}}
\multicolumn{10}{c}{(a) {\bf ASR+SID task} given 8 speaker profiles} \\
    \toprule
System & Latency on ASR & \multirow{2}{*}{\shortstack[c]{ Latency on SID\\(avg. duration)}}& \multicolumn{7}{c}{cpWER (\%) for different overlap ratio} \\ \cmidrule{4-10}
      & &   & 0L & 0S & 10 & 20 & 30 & 40 & {\bf Total}  \\ \midrule % \hdashline[0pt/2pt]\hdashline[0pt/2pt] \midrule
% \multicolumn{10}{l}{\it (Non-streaming Models with VAD)}  \\      
SOT LSTM SA-ASR \cite{kanda2020investigation}             & $\infty^\star$  & $\infty^\star$ & 8.0 & 15.7 & 12.5 & 17.5 & 24.3 & 27.6 & {\bf 18.6} \\
SOT Conformer SA-ASR \cite{kanda2021end}      & $\infty^\star$  & $\infty^\star$ & 7.9 & 12.1 & 9.6 & 10.8 & 13.4 & 15.7 & {\bf 11.9} \\ \midrule
% 30-60 (shrink version)
% SA-TT-18              & 0.16 sec  & + 4.68 sec  & 8.7 & 10.4 & 10.2 & 12.4 & 17.2 & 19.0 & {\bf 13.5} \\
% SA-TT-36              & 0.16 sec  & + 4.68 sec  & 8.2 & 9.2 & 9.6 & 12.3 & 16.1 & 17.9 & {\bf 12.7} \\ % \hdashline[1pt/2pt]\hdashline[0pt/1pt] 
% SA-TT-36              & 2.56 sec  & + 4.68 sec  & 6.0 & 10.6 & 9.8 & 10.6 & 13.8 & 15.6 & {\bf 11.5} \\ \bottomrule
%%% 2.5k model
% SA-TT-18              & 0.16 sec  & 2 words (0.62 sec)  & 10.3  & 11.8 & 12.0 & 14.8 & 20.0 & 20.9 & {\bf 15.6} \\
% SA-TT-36              & 0.16 sec  & 2 words (0.62 sec)  & 9.9 & 12.2 & 11.9 & 14.4 & 18.9 & 20.0 & {\bf 15.1} \\
% SA-TT-36              & 2.56 sec  & 2 words (0.62 sec)  & 6.6 & 12.3  & 10.2 & 12.3 & 14.9 & 15.7 & {\bf 12.5} \\ \bottomrule
%%% 25k model
t-SOT TT-18 w/ t-vector              & 0.16 sec  & + 2 words (0.70 sec)  & 10.3  & 11.8 & 12.1 & 14.6 & 19.5 & 20.2 & {\bf 15.3} \\
t-SOT TT-36 w/ t-vector              & 0.16 sec  & + 2 words (0.70 sec)  & 10.0 & 11.2 & 11.6 & 14.7 & 18.2 & 19.1 & {\bf 14.6} \\
t-SOT TT-36 w/ t-vector              & 2.56 sec  & + 2 words (0.70 sec)  & 6.6 & 11.2  & 9.9 & 11.6 & 14.4 & 15.5 & {\bf 12.0} \\ \bottomrule
\multicolumn{10}{c}{\scriptsize 
 $^\star$ Latency is determined by the segment length based on VAD. The average length of speech segments in LibriCSS is 13.10 sec.  }\\
\vspace{-2mm} \\
\multicolumn{10}{c}{(b) {\bf ASR+SD task} given the oracle number of speakers} \\
    \toprule
System & Latency on ASR & \multirow{2}{*}{\shortstack[c]{ Latency on SD\\(avg. duration)}}& \multicolumn{7}{c}{cpWER (\%) for different overlap ratio.} \\ \cmidrule{4-10}
      & &   & 0L & 0S & 10 & 20 & 30 & 40 & {\bf Total}  \\ \midrule % \hdashline[0pt/2pt]\hdashline[0pt/2pt] \midrule
% \multicolumn{10}{l}{\it (Non-streaming Models with VAD)}  \\      
TS-VAD $\rightarrow$ Transformer ASR \cite{raj2020integration} & $\infty^\diamond$ & $\infty^\diamond$ & 9.5 & 11.0 & 16.1 & 23.1 & 33.8 & 40.9 & {\bf 23.9} \\
SOT LSTM SA-ASR $\rightarrow$ NME-SC \cite{kanda2020investigation}             & $\infty^\diamond$  & $\infty^\diamond$ & 10.3 & 15.8 & 13.4 & 17.1 & 24.4 & 28.6 & {\bf 19.2} \\
SOT Conformer SA-ASR$\rightarrow$ NME-SC \cite{kanda2021end}      & $\infty^\diamond$  & $\infty^\diamond$ & 8.6 & 12.7 & 11.2 & 11.3 & 16.1 & 17.5 & {\bf 13.3} \\ % \midrule
NME-SC $\rightarrow$ SOT Conformer SA-ASR \cite{kanda2021transcribe}      & $\infty^\diamond$  & $\infty^\diamond$ & 9.0  & 12.2 & 8.7 & 10.9 & 13.7 & 13.9 & {\bf 11.6} \\ \midrule
% \multicolumn{10}{l}{\it (Streaming Models)}  \\      
%% 30-60, th0.98, with large delay DvecExtractorNet
% SA-TT-18            & 0.16 sec  & + 4.68 sec  & 11.6 & 11.4 & 10.6 & 12.8 & 17.9 & 19.6 & {\bf 14.4} \\ 
% SA-TT-36            & 0.16 sec  & + 4.68 sec  & 8.5  & 11.0 & 9.7 & 12.5 & 15.8 & 17.7 & {\bf 13.0} \\ 
% SA-TT-36            & 2.56 sec  & + 4.68 sec  & 5.8 & 9.9 &  12.1 & 11.2 & 14.9 & 16.7 & {\bf 12.3} \\  \bottomrule
%%% 2.5k model
% SA-TT-18            & 0.16 sec  & 2 words (0.62 sec)  & 8.9 & 12.5 & 12.4 & 15.3  & 21.3 & 21.8  & {\bf 16.1} \\ 
% SA-TT-36            & 0.16 sec  & 2 words (0.62 sec)  & 11.6 & 11.8 & 11.4 & 15.5 & 19.5 & 22.2 & {\bf 15.9} \\ 
% SA-TT-36            & 2.56 sec  & 2 words (0.62 sec)  & 6.4 & 11.5 & 11.5 & 12.5 & 17.3 & 17.8 & {\bf 13.4} \\ \bottomrule 
%%% 25k model
t-SOT TT-18 w/ t-vector            & 0.16 sec  & + 2 words (0.70 sec)  & 9.0 & 13.5 & 12.1 & 15.4  & 19.4 & 21.4  & {\bf 15.8} \\ 
t-SOT TT-36 w/ t-vector            & 0.16 sec  & + 2 words (0.70 sec)  & 9.1 & 10.8 & 11.9 & 14.2 & 18.3 & 20.4 & {\bf 14.7} \\ 
t-SOT TT-36 w/ t-vector            & 2.56 sec  & + 2 words (0.70 sec)  & 6.8 & 10.2 & 10.5 & 11.6 & 15.4 & 16.7 & {\bf 12.4} \\ \bottomrule 
\multicolumn{10}{c}{\scriptsize %\footnotesize 
 $^\diamond$ Latency is determined by the length of the recording. The average length of the LibriCSS recordings is 10 min.  }
  \end{tabular}
  }
  \vspace{-6mm}
\end{table*}

\begin{table}[t]
% \ra{1.0}
\ra{0.93}
  \caption{cpWER (\%) for LibriCSS based on SA-TT-36 (2.56 sec ASR latency) with different delayed decisions on SID/SD.}
  \label{tab:libricss_latency}
  \vspace{-3mm}
  \centering
 { \scriptsize
%  {  \footnotesize
  % \tabcolsep = 2.0mm
\begin{tabular}{@{}llrrrrr@{}}
    \toprule
Model & Task &  \multicolumn{5}{c}{\# of delayed words on SID/SD}  \\ \cmidrule{3-7}
      & &  0 & 1 & 2 & 4 & 8  \\ \midrule 
% 2.5k model, th=0.98
% SA-TT-18  & SID & 16.7 & 15.9 & 15.6 & {\bf 15.4} & 18.5 \\ 
% SA-TT-18  & SD  & 17.0 & {\bf 16.1} & {\bf 16.1} & 17.7 & 21.6 \\ \bottomrule
% 25k model, th=0.98
t-SOT TT-36 w/ t-vector  & ASR+SID & 12.6 & 12.1 & {\bf 12.0} & 12.2 & 15.5 \\ 
t-SOT TT-36 w/ t-vector  & ASR+SD  & 13.1 & 12.7 & {\bf 12.4} & 14.3 & 19.7  \\ \bottomrule
  \end{tabular}
  }
  \vspace{-5mm}
\end{table}

\section{Experiments}
\label{sec:experiments}
We first conducted a preliminary experiment for ASR and SID task
by using LibriSpeechMix \cite{kanda2020sot},
where the number of speakers in each audio is limited to two.
We then extended the experiment by using LibriCSS \cite{chen2020continuous}
to evaluate the
SA-ASR performance, with both SID and SD, 
for long-form audio containing many utterances.

\subsection{Experiment with LibriSpeechMix}
\subsubsection{Experimental settings}
\label{sec:librispeechmix_setting}
We used the single speaker and two speaker subsets of
LibriSpeechMix \cite{kanda2020joint} as
a preliminary experiment for ASR and SID task. % with up to two speakers.
The dataset is
 made by mixing up to two utterances randomly sampled from LibriSpeech \cite{panayotov2015librispeech}
with a randomly-determined delay.
% such that
% each evaluation sample contains a partial speaker overlaps.
For each test sample, %8 speaker profiles are provided, 
 an SA-ASR model/system must 
generate a multi-talker transcription while
 identifying the speaker of each word from 8 candidate speakers given
 their speech profiles,
which were extracted by ResNet-based d-vector \cite{gao2019res2net,xiao2021microsoft}. % (2 utterances for each candidate speaker).
We used SER, WER and speaker-attributed WER (SAWER), 
as defined in \cite{kanda2020joint}, 
as our evaluation metrics. 
Here, SAWER is our primary metric, 
where the estimated word
is regarded as correct if and only if an estimated word and speaker 
both match the reference.

For the ASR block, we used a 18-layer or 36-layer TT (TT-18 or TT-36 in short)
with the chunk-wise look-ahead proposed in \cite{chen2021developing}
by using exactly the same configuration as in \cite{kanda22arxiv}.
Each transformer block consisted of
a 512-dim MHA with 8 heads 
and a 2048-dim point-wise feedforward layer.
The prediction network consisted of
2 layers of 1024-dim LSTM.
4,000 word pieces plus blank and $\langle \mathrm{cc}\rangle$ tokens
were used
as the recognition units.
The input feature is an 80-dim log mel-filterbank extracted every 10 msec.
Refer to \cite{kanda22arxiv} for further details on the ASR block.
For the proposed speaker embedding extraction, 
 Res2Net \cite{gao2019res2net,xiao2021microsoft} was used for ${\rm DvecExtractorNet}$ 
and a 128-dim 8-head MHA was used at each layer of the speaker encoder.
The speaker decoder consists of a 2-layer 512-dim LSTM.
As proposed in \cite{chen2021developing}, 
we controlled the algorithmic latency
of the model
based on the chunk size of the attention mask.
%For the remainder of the paper, 
%We call
%  the 18-layer and 36-layer models as SA-TT-18 and SA-TT-36, respectively.

For the model training, 
we simulated the training data by randomly mixing up to two utterances
from ``train\_960'' of LibriSpeech by following the procedure 
in \cite{kanda22arxiv}.
For each training sample, we randomly selected up to 8 candidate speakers 
to calculate $\mathcal{L}^{\rm spk}$.
We initialized the ASR block by using the parameters of the t-SOT ASR
 trained in our prior work \cite{kanda22arxiv}, 
and the ASR block was frozen during the training. % with $\mathcal{L}^{\rm spk}$.
We performed 60K training iterations with
16 GPUs, each of which consumed mini-batches of 12,000 frames.
We used an AdamW optimizer with
a linear decay learning rate schedule with a peak learning rate 
of 1.5e-3 
after 10K warm up iterations.

\subsubsection{Evaluation results}
\label{sec:librispeechmix_result}
The evaluation 
results are shown in Table \ref{tab:librispeechmix_summary}.
We first observed that the proposed t-SOT TT-18 with t-vector-based SID
significantly outperformed the prior streaming model, SURIT \cite{lu2021surit},
on both SER and WER even with a SID latency of only 2 words  (0.68 sec on average).
With a longer SID delay of 8 words, which is still faster than SURIT, all metrics were further
improved.
We then observed that 
both the model size 
and
the latency on ASR 
 had a significant impact on WER and SAWER.
The model based on TT-36 with 2.56 sec of ASR latency even achieved a comparable
SAWER for the single speaker test set and a better SAWER for the two-speaker-mixed test set
than the prior offline models.

\subsection{Experiment with LibriCSS}
\subsubsection{Experimental settings}
To evaluate the proposed model in a more realistic setting,
we conducted an evaluation with LibriCSS \cite{chen2020continuous}.
LibriCSS
 is a set of 10-min
recordings (10 hours in total), created by playing back 
utterances of  % in ``test\_clean''
 LibriSpeech ``test\_clean'' in a real meeting room.
The original recording was made with a 7-ch microphone array,
and the first channel of the recording was used
in our experiment.
 Each 10-min recording are made of utterances from 8 speakers
 and categorized by the speaker overlap ratio from 0\% to 40\%.
 We evaluated the proposed model based on
concatenated minimum-permutation WER (cpWER) \cite{watanabe2020chime},
 which is affected by both ASR and speaker attribution errors.

For training the model,
we simulated the training data by following 
the procedure of \cite{kanda22arxiv},
where each training audio consists of up to 5 utterances 
randomly sampled from LibriSpeech ``train\_960''.
Up to 8 candidate speakers were randomly selected 
from LibriSpeech ``train\_960''
to compute $\mathcal{L}^{\rm spk}$.
We initialized the ASR block with the parameters
of t-SOT TT trained for LibriCSS in \cite{kanda22arxiv}, and the speaker block 
with the parameters of the model trained for LibriSpeechMix.
The ASR block was frozen during the training using $\mathcal{L}^{\rm spk}$.
We performed 25,000 training iterations with
16 GPUs, each of which consumed a mini-batch of 12,000 frames.
The AdamW optimizer was used with
a linear decay learning rate schedule  
with a peak learning rate of 1.5e-4.

In the inference, % with our proposed model, 
we stopped the beam search
every time a silence was detected by WebRTC Voice Activity Detector
\cite{webrtcvad}
after 20 seconds 
of decoding. We forcibly terminated the beam search when a silence was not detected
after 40 seconds of decoding.
For the ASR and SID task, we supplied 8 speaker profiles corresponding to
the speakers in the recordings, where each speaker profile is extracted
by using 5 utterances of the speaker.
For the ASR and SD task, we used 
 the normalized maximum eigengap-based
spectral clustering (NME-SC) \cite{park2019auto} given
the oracle number of speakers.
The threshold for detecting the speaker change was set to 0.98.

\subsubsection{Evaluation results}

Table \ref{tab:libricss_summary} shows the main result. 
Note that, 
this is the first time that 
a fully streaming SA-ASR system
was applied to LibriCSS, and thus
we don't have the numbers for the prior streaming models in the table.
The proposed t-SOT TT model with t-vector achieved 
cpWERs of 12.0\% and 12.4\% for ASR+SID and ASR+SD, respectively,
which are both very close to the SOTA results of the offline SA-ASR model.
% For ASR and SID task,
% the proposed SA-TT model achieved a cpWER of 12.0\%, 
% which is on par to 
% the cpWER based on the SOTA offline SA-ASR model.
% For SD task, the proposed SA-TT model achieved a cpWER
% of 12.4\%,
% which is close to 
%  the result based on the prior offline model (11.6\%).

Table \ref{tab:libricss_latency}
shows the effect of the delayed decisions on SID/SD.
As discussed in Section \ref{sec:streaming-sid},
a longer delay stabilizes the speaker embedding at the cost of
increased miss errors of speaker change points.
% As observed in LibriSpeechMix (Section \ref{sec:librispeechmix_result}),
% generally, the longer the delay is, the better the estimated embedding becomes.
% However, the longer the delay is, the higher the chance of missing speaker change 
% detection becomes.
Unlike LibriSpeechMix which contains up to 2 utterances,
the drawback of missing speaker change points 
had a larger impact for LibriCSS,
%  which contains many utterances,
%which 
 resulting in a significant degradation of cpWER
with the 4- and 8-word delays.

\section{Conclusion}
This paper presented % SA-TT,
a new streaming SA-ASR model.
We proposed
an encoder-decoder based speaker embedding extractor 
that synchronously works with the t-SOT ASR,
and showed a simple streaming SID/SD procedure with estimated speaker embeddings, t-vector.
In our
evaluation, % by using LibriSpeechMix and LibriCSS corpora, 
we showed that the proposed model achieved
comparable or sometimes even superior SA-ASR accuracy to
the SOTA offline models
while enabling  low-latency inference.

\bibliographystyle{IEEEtran}

\bibliography{mybib}

\end{document}